\documentstyle[aps,prl,epsf,twocolumn]{revtex}

\begin{document}

\title{Enhancement of Piezoelectricity in a Mixed Ferroelectric}

\author{Eric Cockayne and Karin M. Rabe}

\address{Department of Applied Physics,
Yale University,
P.O. Box 208284,
New Haven, CT 06520-8284}
\date{\today}
\maketitle

\begin{abstract}

 	We use first-principles density-functional
total energy and polarization calculations to
calculate the piezoelectric tensor at zero temperature
for both cubic and simple tetragonal ordered supercells
of Pb$_3$GeTe$_4$.
The largest piezoelectric coefficient
for the tetragonal configuration is enhanced by a factor
of about three with respect to that of the cubic 
configuration.  This can be attributed to both the larger
strain-induced motion of cations relative to anions
and higher Born effective charges in the tetragonal case.  
A normal mode decomposition shows that both cation
ordering and local relaxation weaken the ferroelectric
instability, enhancing piezoelectricity.

\end{abstract}


  Piezoelectric materials are of great technological
importance.  The highest known piezoelectric coefficients are found
in mixed compounds, such as Pb(Zr$_{1-x}$Ti$_x$)O$_3$ near
its morphotrophic phase boundary at
$x \approx 0.45$. 
This association has been reinforced by the recent discovery of
giant piezoelectricity in
single crystals of the relaxor 
ferroelectric systems
Pb(A$_{1/3}$Nb$_{2/3}$)O$_3$--PbTiO$_3$, 
(A = Zn, Mg).\cite{Par96}
Understanding of the physical origin of enhancement
of piezoelectricity in mixed ferroelectrics would be of great
theoretical
interest and could also point to new ways to tune this property.

  Ab initio calculations have proved successful in relating 
properties of ferroelectrics to phenomena on the atomic 
level.
Structural parameters\cite{Coh92,Sin95}, 
dielectric constants, effective charges and
phonon dispersion relations\cite{Bar87,Gon92}, 
and structural phase 
transition temperatures\cite{Rab87} have been
calculated from first principles.
Such methods can naturally be extended to piezoelectricity.
Piezoelectric coefficients have been calculated from first
principles for various materials\cite{deG89,Dal94,Sag97},
where good agreement with experimental results
is obtained.
For mixed systems,
{\it ab initio} 
calculations can be performed on individual ordered 
supercells and the results analyzed to understand
the dependence on configuration and the
resulting ensemble average. 
In this Letter, we consider the mixed ferroelectric
Pb$_{0.75}$Ge$_{0.25}$Te.
Calculation of the piezoelectric response shows significant
configuration dependence, with
enhancement of piezoelectricity in the cell with lower
symmetry.
We interpret and explain the
results in terms of chemical bonding
and long range dipolar interactions.

\begin{figure}[htbp]
  \begin{center}
    \leavevmode
    \epsfxsize = 3.3in
    \epsfbox{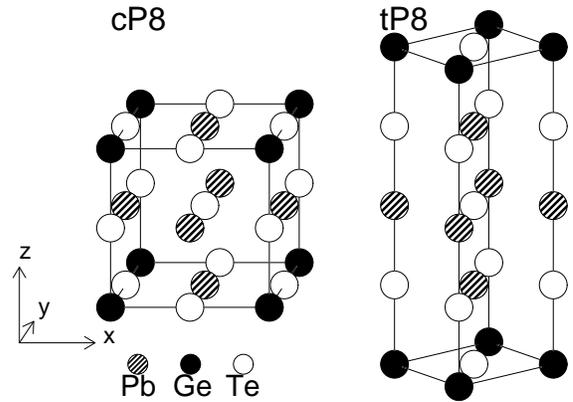}
  \end{center}
  \caption{Pb$_3$GeTe$_4$ cP8 and tP8 configurations.}
  \label{both.fig}
\end{figure}

  Pb$_{1-x}$Ge$_x$Te undergoes a transition from a
paraelectric cubic rocksalt phase at high temperature to a 
ferroelectric rhombohedral phase at low temperature, 
dominated by Ge
off-centering\cite{Log77},
for all compositions $x > 0.005$.\cite{Tak79}.  
In order to study the dilute Ge limit while maintaining
small supercells,
we choose
$x = 0.25$.
The configurations which we investigate are shown in 
Figure~\ref{both.fig}.   The structures
have identical compositions Pb$_3$GeTe$_4$, but one is
simple cubic (cP8) and one is simple tetragonal (tP8).  

  First-principles density-functional calculations 
were performed using the
program CASTEP 2.1\cite{Cas91}, 
with a plane wave basis set in
the local density approximation.  
Bachelet-Hamann-Schl\"uter pseudopotentials\cite{Bac82} 
were
used for all atoms.  For the cP8 configuration, we
used a $4 \times 4 \times 4$ Monkhorst-Pack set of 
${\bf k}$ points; for the tP8 configuration, we used
the same ${\bf k}$ point grid, folded into the tetragonal
Brillouin zone.
In our study of the structural
phase transition in the
Pb$_3$GeTe$_4$-cP8 configuration\cite{Coc97},
we give further details on the {\it ab
initio} calculations. 
Fixing the atoms in rocksalt
positions and minimizing total energies with
respect to strain, we obtained
$a = 6.275$~\AA~for cP8 and $a = 4.438$~\AA; 
$c = 12.548$~\AA~for tP8.
These are taken as the high symmetry reference
structures for the two configurations.
Next, we determined the ground state structures.
For a series of fixed global strains, as many as 6 
independent 
internal coordinates in the cP8 configuration and 16 
internal
coordinates in the tP8 configuration 
were relaxed via conjugate gradients
minimization.
In each relaxation, we broke the initial symmetry
by displacing
each Ge atom by 0.3~\AA~in the (111) direction.
A quadratic fit of total energies as a
function of strain  was used to find the
equilibrium strain.
The
cP8 ground state obtained is
rhombohedral, space
group R3m, $a = 6.315$ \AA, $\alpha = 89.47^{\circ}$,
and the tP8 ground
state is monoclinic,
space group Pm, $a = 4.497$~\AA,
$b = 4.409$~\AA, $c = 12.575$~\AA, $\beta = 89.31^{\circ}$.
The internal coordinates are given in Table~\ref{ground.tbl},
with the center of mass fixed in each case to that
of a strained rocksalt structure with Ge on the origin.

  The method of King-Smith and Vanderbilt (KSV)\cite{Kin93} 
was then used to calculate polarizations. 
This method calculates the {\it electronic} contribution
to the polarization, to which the ionic contribution is
added to obtain the total polarization $\bf P$.
The component of polarization along lattice vector
${\bf a}_{\alpha}$ is expressed in terms of
a set of
wavefunction calculations along 
straight-line
paths in the 
Brillouin zone, where 
reciprocal lattice vector ${\bf b}_{\alpha}$ is the path vector,
$J$ the number of $\bf k$ points in the path, and 
${\bf k}_{\perp}$ the component of ${\bf k}$ orthogonal 
to ${\bf b}_{\alpha}$ along the path.

  In Voigt notation, the components of the
(third rank) piezoelectric strain tensor
are given by 
$e_{\alpha\beta} \equiv \partial P_{\alpha}/ \partial
\epsilon_{\beta}$,
where $\epsilon_{\alpha}$ is a
component of the (second rank) strain tensor.
We used finite differences to determine the
piezoelectric tensor components:
$e_{\alpha\beta} \approx \Delta P_{\alpha}/
\Delta \epsilon_{\beta}$. 
For both the cP8 and tP8 configurations, we
calculated
${\bf P}$ for the ground states and
for a set of fully relaxed strained cells 
with one strain component $\epsilon_{\alpha}$ 
increased by 0.005.  Strains were measured 
with respect to the axes shown in Figure~\ref{both.fig}.  
Strain magnitudes were measured with respect to the
cP8 reference structure ($a = 6.275$~\AA) in both
cases.  
The KSV calculations were performed using the
self-consistent potentials obtained from total
energy calculations using 64 ${\bf k}$ points in the Brillouin
Zone.
We used $J = 32$
for cP8, $J = 16$ for tP8 along ${\bf b}_3$, and
$J = 48$ for tP8 along ${\bf b}_1$ and ${\bf b}_2$.
In each case, we obtained the cross section
average by averaging the values for ${\bf k}_{\perp}$ on a 
$4 \times 4$ Monkhorst-Pack grid.  These parameters
were sufficient to
converge the
piezoelectric tensor components to within 0.05
$C/m^2$.  

   The resultant piezoelectric tensor components
and their decomposition into clamped ion and internal
strain
contributions\cite{Dal94}
are shown in Table~\ref{compare.tbl}.  
Generally, the components in the tP8
configuration are larger in magnitude than in the cP8 one.
The clamped ion terms are remarkably similar; 
the differences are almost entirely due to internal
strain.
We will focus on $e_{33}$ because it 
is the largest piezoelectric component in each case
and shows the greatest change in going from the
cP8 to the tP8 configuration.

  We first look at the configuration dependence
of the internal strain term $e_{33}|_{\rm int}$
by analyzing the contributions of individual
ions:
\begin{equation}
\label{contrib.eqn}
e_{33}|_{\rm int} = \frac{1}{V}
\sum_{i \alpha} {\bf Z}_{i, z \alpha}^{\star} 
{{\partial  u_{i \alpha}}\over
{\partial \epsilon_3}},
\end{equation}
where ${\bf Z}_i^{\star}$ are the Born effective charge
tensors and
$\partial {\bf u}_i / \partial \epsilon_{3}$ is the
movement of the relaxed ion position from the
clamped ion position with strain (in fixed 
center of mass
coordinates).
For each configuration,
we determined
${\bf Z}^{\star}_i$
from the changes in polarization
resulting from displacing each atom in
turn
0.01~\AA~along each axis direction,
and $\partial {\bf u}_i/ \partial \epsilon_3$
by comparing the relaxed coordinates of the ions in the
two cells differing only in $\epsilon_3$.
Since the displacement derivatives
$\partial {\bf u}_i/ \partial \epsilon_3$
are dominated by motion in the $\hat z$ direction,
we make the following approximation:
\begin{equation}
\label{quasi1d.eqn}
e_{33}|_{\rm int} \approx \tilde{e}_{33}|_{\rm int} \equiv \frac{1}{V}
\sum_i Z^{\star}_{i,zz} {{\partial u_{iz}}\over
{\partial \epsilon_{3}}} \equiv 
\sum_i^{\rm unit~cell} \tilde{e}_{33}^{(i)}|_{\rm int}
\end{equation}
The calculated values for $\partial u_{iz} / \partial \epsilon_3$
and $Z^{\star}_{i,zz}$ are given for each ion in Table \ref{source.tbl}.

 For both the cP8 and tP8 configurations,
  the values of $\tilde{e}_{33}|_{\rm int}$
obtained from Eq.~\ref{quasi1d.eqn}
agree with ${e}_{33}|_{\rm int}$
to better than 6\%.
The values of Table~\ref{source.tbl} can thus be used
to associate the enhancement of piezoelectricity in tP8
with the response of the individual ions.
It can be seen that the increased ionic motion under strain
is the dominant factor, 
with additional enhancement due
to the larger average magnitude of the Born effective
charges.
Both the magnitude of the piezoelectric response and
the configuration dependence is greatest in
the chains along $\hat z$ containing the Ge ions.

  The theory of resonant p-bonding \cite{Luc73,Lit84} 
in IV-VI compounds gives insight into this trend.
In the symmetric rocksalt structure, the covalent bonds are
composed primarily of p orbitals, with 
one electron per bond.  In each anion-cation-anion
segment, there is thus the tendency
for the cation to prefer an off-center position, forming
a short cation-anion bond on one side into which the electron
from the long bond is partially transferred.
The off-centering is therefore associated with a large ${\bf Z}^{\star}$.
This tendency is counterbalanced by short range
repulsion, the strength of which is roughly determined by the
ionic radius of the cation.
In pure PbTe, short-range repulsion dominates and the low temperature
structure is rocksalt\cite{Lit84}, while in GeTe, the
off-centering instability dominates and the
low temperature structure is rhombohedral.
In Pb$_{0.985}$Ge$_{0.015}$, the low temperature structure
is rhombohedral at low pressure and cubic at high pressure\cite{Sus83},
showing that the relative importance of the short-range 
repulsion increases with pressure.

\begin{figure}[htbp]
  \begin{center}
    \leavevmode
    \epsfxsize = 3.3in
    \epsfbox{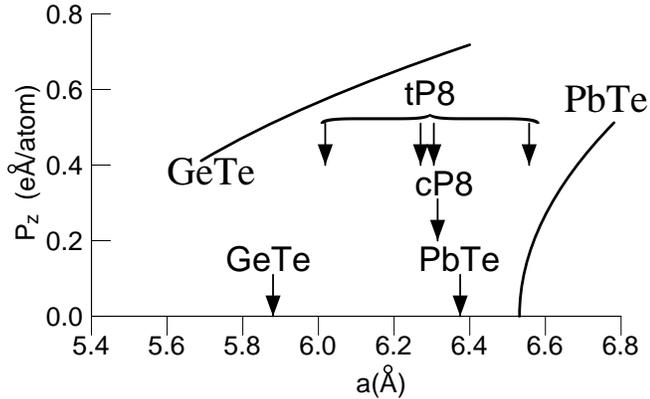}
  \end{center}
  \caption{Polarization vs. lattice parameter in PbTe
and GeTe and first-principles LDA
``local lattice parameters" along
$\hat z$ for PbTe, GeTe and mixed Pb$_3$GeTe$_4$
configurations}
  \label{pola.fig}
\end{figure}

Equivalently, a decrease in the Te-cation-Te distance
suppresses off-centering.
Figure~\ref{pola.fig} shows the $z$-component of the
polarization of pure GeTe and PbTe as a function of lattice
parameter, fit to the results of KSV calculations in which 
internal relaxation is allowed but the lattice is held cubic.
A ferroelectric instability is predicted for PbTe at
sufficient negative pressure.
In the mixed system, we define {\it local} lattice
parameters along $\hat z$ to be equal to the Te-cation-Te
distance,
and qualitatively associate the local piezoelectric 
response with the
slope of the $P_z$ vs. $a$ curve for the
corresponding pure cation-Te system.
In the cP8 configuration, there is a single local lattice 
parameter, the value of which favors off-centering and
piezoelectric response for the Te-Ge-Te chain but not for
the Te-Pb-Te chains.  In the tP8 configuration, there is
a different type of chain, with alternating Ge and Pb.
In this chain, relaxation of the Te ions along $\hat z$
leads to two effective local lattice constants: a 
short Te-Ge-Te and a large Te-Pb-Te distance marked in
Figure~\ref{pola.fig}.  The latter Te-Pb-Te distance is
considerably larger than in pure PbTe, favoring
off-centering and piezoelectric response in the Te-Pb-Te
segment as well as the Te-Ge-Te segment, as seen in 
Table~\ref{source.tbl}.

  Another informative way to analyze the piezoelectric
response is to decompose the ionic displacements into
the contributions of the zone-center normal modes of the
system.
For the decomposition of $\tilde e_{33}|_{int}$, we need only
the polar modes of the high symmetry reference structures
involving ionic motion along $\hat z$.
For the tP8 systems, two reference structures 
were considered.
One, tP8$_{un}$, has all ions sitting on ideal rocksalt
coordinates,
the other, tP8$_{rel}$, is derived from
tP8$_{un}$ via
a {\it symmetry preserving} relaxation.
In this
relaxation, the Te atoms in the $ z = \pm 0.25$ plane move
$\mp 0.143 \hat z$~\AA~and the Pb atoms in the $ z = \pm 0.25$ plane move
$\mp 0.029 \hat z$~\AA.
The $\hat z$ polar modes were determined from
first principles frozen phonon calculations, as in Ref.
\cite{Coc97}, with the results given in Table~\ref{pmode.tbl}.
For each reference structure, one mode has imaginary frequency,
indicating instability to a symmetry-breaking distortion.

  The approximate piezoelectric
response $\tilde e^{33}|_{int}$ for a given ground state can be 
decomposed into 
normal mode contributions:
\begin{equation}
\tilde e_{33}|_{int} = \frac{1}{V} \sum_{\mu} \overline{Z}^{\star}_{\mu}
\frac{\partial \xi_{\mu}}{\partial \epsilon_3},
\end{equation}
where mode effective charges are given by
$\overline{Z}^\star = \sum_i^{\rm unit~cell}
Z^{\star}_{i,zz} \left( \partial u_{iz}/
\partial \xi_{\mu} \right)$ and
${\partial \xi_{\mu}}/{\partial \epsilon_3}$ is
the change in the amplitude of mode $\mu$ with strain.
All modes are normalized so that the
sum of the squared atomic displacements is 1 \AA$^2$ per
8-atom cell.

The results are shown in Table \ref{pmode.tbl}.
In each case, the largest mode contribution is from the unstable mode.
In fact, the unstable tP8$_{rel}$ mode accounts
for 97\% of $\tilde e^{33}|_{int}$ for tP8.
We find that the relative Pb(3) motion in the unstable
mode goes from 0.16 for tP8$_{un}$ to 0.49 for tP8$_{rel}$,
in agreement with the local viewpoint that a sufficient
increase in the Te-Pb-Te distance will favor Pb off-centering
and piezoelectric response.

 We investigated the unstable mode contribution in greater
depth by constructing a simple ``double well" parametrization
of the energy as a function of the amplitude of unstable
mode distortion and strain $\epsilon_3$.
When all other parameters are equal, decreasing the strength of
the quadratic instability increases the piezoelectric 
response associated with the unstable mode.
This trend is clear in Table~\ref{pmode.tbl}.
The weakening of the tP8$_{un}$ instability relative to the
cP8 instability is the result of the change in 
cationic configuration and is due largely to less favorable
dipole-dipole interactions.
Symmetry-preserving relaxation further weakens the
instability and leads to an even greater unstable mode
response.

 As $\nu^2 \rightarrow 0^-$ for an unstable mode,
the associated piezoelectric response will {\it diverge}.
In stoichiometric ferroelectrics, this is essentially
what happens at the paraelectric-ferroelectric
transition, with a corresponding divergence in the
piezoelectric coefficients on the low-symmetry side.
In substitutionally disordered compounds containing one stable and one unstable
endmember, such as Pb$_{1-x}$Ge$_{x}$Te, there will be
a variety of local environments and therefore a 
range in the degree of local (in)stabilities.
We expect in such systems that enhanced 
piezoelectricity can be observed for a range of
temperatures, pressures and applied fields where
the density of marginally unstable optical modes is significant.
Even at fixed composition, the type of ordering present
has a dramatic effect on the local stability and could
be  used to tune the piezoelectricity,
({\it e.g.} through control of processing
or in epitaxial thin films).
Quantitative modeling of the temperature-dependent response
associated with the unstable modes of particular
configuration, using the
effective Hamiltonian approach \cite{Rab95lwf}, is
currently in progress.

In conclusion, we have compared the piezoelectricity in two
Pb$_3$GeTe$_4$ configurations and found strong enhancement of
$e_{33}$ in the tetragonal configuration.
This effect is intrinsic; it will occur even
in a single grain, single phase sample.
The interpretation of the calculated results in terms of 
local polar instabilities and the long-range interactions
between them suggests further opportunities for identification
of solid solutions with large piezoelectric response.

  This work was supported by ONR  N00014-97-J-0047.
We thank Volker Heine for stimulating discussions.

\baselineskip = 2\baselineskip  

\newpage

\begin{table}[e]
\caption{{\it Ab initio} ground state structures of
Pb$_3$GeTe$_4$ cP8 and tP8 configurations.  Lattice constants
given in text.}
\begin{tabular}{||c|c|c|
c||c|c|c|c||}
\multicolumn{4}{c||} {cP8}  &
\multicolumn{4}{c||} {tP8}  \\
\hline Atom & $x$  & $y$ & $z$ & Atom & $x$ & $y$ & $z$ \\
\hline
Pb(1)  & 0.0014 & 0.5037 & 0.5037 & Pb(1) & 0.5120 & 0.5  & 0.2485 \\
Pb(2)  & 0.5037 & 0.0014 & 0.5037 & Pb(2) & 0.4961 & 0.5  & 0.7516 \\
Pb(3)  & 0.5037 & 0.5037 & 0.0014 & Pb(3) & 0.0054 & 0    & 0.5065 \\
Ge  & 0.0308 & 0.0308 & 0.0308 & Ge & 0.0576 & 0    & 0.0062 \\
Te(1)  & 0.9985 & 0.9985 & 0.4760 & Te(1) & 0.9985 & 0    & 0.2349 \\
Te(2)  & 0.9985 & 0.4760 & 0.9985 & Te(2) & 0.9849 & 0    & 0.7563 \\
Te(3)  & 0.4760 & 0.9985 & 0.9985 & Te(3) & 0.4661 & 0.5  & 0.9980 \\
Te(4)  & 0.4955 & 0.4955 & 0.4955 & Te(4) & 0.4958 & 0.5  & 0.4966 \\
\end{tabular}
\label{ground.tbl}
\end{table}

\begin{table}[e]
\caption{Comparison of piezoelectric tensors for
the ground states of two Pb$_3$GeTe$_4$
configurations
(in $C/m^2$).  Components in parentheses are equal to other
components via symmetry.}
\begin{tabular}{||c|c|c|
c|c|c|c||}
Component  & \multicolumn{2}{c|} {Total}  &
\multicolumn{2}{c|} {Clamped ion}  &
\multicolumn{2}{c||} {Internal strain} \\
\hline          &  cP8  & tP8   & cP8   & tP8   &  cP8  &  tP8  \\
\hline $e_{11}$ &   1.8  &  2.5 & -0.5 & -0.5 &  2.3 &  3.0 \\
\hline $e_{12}$ &  -0.6  & -1.2 & 0.0 & 0.0 &  -0.6 &  -1.2 \\
\hline $e_{13}$ & (-0.6) & -0.9 & (0.0) & 0.0 & (-0.6)  & -0.9 \\
\hline $e_{14}$ &  -0.2  & -0.6 & 0.0 & 0.0 &  -0.2 &  -0.6 \\
\hline $e_{15}$ &   1.0  &  1.2 & 0.6 & 0.5 &  0.4 &  0.7 \\
\hline $e_{16}$ &  (1.0) &  1.1 & (0.6) & 0.5 &  (0.4) &  0.6 \\
\hline $e_{31}$ & (-0.6) & -0.3 & (0.0) & 0.0 &  (-0.6) &  -0.3 \\
\hline $e_{33}$ &  (1.8) &  5.1 & (-0.5)  & -0.5 &  2.3 &  5.6 \\
\hline $e_{34}$ &  (1.0) &  2.2 & (0.6) & 0.7 &  (0.4) &  1.5 \\
\hline $e_{36}$ & (-0.2) & -0.5 & (0.0) & 0.0 &  (-0.2) &  -0.5 \\
\end{tabular}
\label{compare.tbl}
\end{table}

\begin{table}[e]
\caption{Contributions of individual ions to
$e_{33}|_{\rm int}$ in
Pb$_3$GeTe$_4$ cP8 and tP8 configurations.
Atoms located in chains along $\hat z$ containing
Ge are italicized.
$\tilde{e}_{33}^{(i)}|_{\rm int}$ are in $C/m^2$;
$\frac{\partial u_{iz}}{\partial \epsilon_3}$ in
\AA.
}
\begin{tabular}{||c|c|c|
c|c|c|c|c||}
\multicolumn{4}{c|} {cP8}  &
\multicolumn{4}{c||} {tP8}  \\
\hline Atom & $Z_{i,zz}^{\star}$ &
$\frac{\partial u_{iz}}{\partial \epsilon_3}$  &
$\tilde{e}_{33}^{(i)}|_{\rm int}$  &
Atom & $Z_{i,zz}^{\star}$ &
$\frac{\partial u_{iz}}{\partial \epsilon_3}$  &
$\tilde{e}_{33}^{(i)}|_{\rm int}$  \\
\hline
Pb(1)        &  5.4 &  0.72 & 0.25 & Pb(1)        &  5.9 &  0.32 & 0.12 \\
Pb(2)        &  5.4 &  0.72 & 0.25 & Pb(2)        &  5.8 &  0.91 & 0.34 \\
Pb(3)        &  5.9 &  0.91 & 0.34 & {\it Pb(3)}  &  5.6 &  3.08 & 1.11 \\
{\it Ge}     &  3.1 &  0.79 & 0.16 & {\it Ge}     &  5.7 &  2.60 & 0.95 \\
{\it Te(1)}  &--4.0 &--3.08 & 0.78 & {\it Te(1)}  &--5.9 &--3.11 & 1.18 \\
Te(2)        &--5.0 &--0.37 & 0.12 & {\it Te(2)}  &--5.7 &--2.80 & 1.03 \\
Te(3)        &--5.0 &--0.37 & 0.12 & Te(3)        &--5.5 &--0.83 & 0.29 \\
Te(4)        &--5.8 &--0.46 & 0.17 & Te(4)        &--5.9 &--1.74 & 0.66 \\
\end{tabular}
\label{source.tbl}
\end{table}

\begin{table*}[e]
\caption{Polar mode contribution to $\tilde{e}_{33}|_{int}$ 
in Pb$_3$GeTe$_4$-cP8 and tP8.  cP8 and tP8 modes have symmetry
labels $\Gamma_{15}^z$ and $\Gamma_{1}^{\prime}$, respectively.
$\overline{Z}^{\star}_{\mu}$ are in e\,\AA;
$\tilde{e}_{33}|_{int}^{(\mu)}$ in $C/m^2$ and $\nu_{\mu}$ in cm$^{-1}$.}
\begin{tabular}{||c|c|c|c||c|c|c|c||c|c|c|c||}
\multicolumn{4}{c||} {cP8}  &
\multicolumn{4}{c||} {tP8$_{un}$}  &
\multicolumn{4}{c||} {tP8$_{rel}$} \\
\hline
$\nu_{\mu}$ & $\overline{Z}^{\star}_{\mu}$ &
${\partial \xi_{\mu}}/{\partial \epsilon_3}$ &
$\tilde e_{33}|_{int}^{(\mu)}$ &
$\nu_{\mu}$ & $\overline{Z}^{\star}_{\mu}$ &
${\partial \xi_{\mu}}/{\partial \epsilon_3}$ &
$\tilde e_{33}|_{int}^{(\mu)}$ &
$\nu_{\mu}$ & $\overline{Z}^{\star}_{\mu}$ &
${\partial \xi_{\mu}}/{\partial \epsilon_3}$ &
$\tilde e_{33}|_{int}^{(\mu)}$ \\
103 i & 5.9 &   2.93  & 1.09 &
64 i & 11.5 & 6.28  & 4.64 &
40 i & 13.6 & 6.32  & 5.53 \\
31 & 1.6 &    1.09  & 0.11 &
42 & 2.9 &    -0.14  &-0.03 &
42 & 5.2 &    0.64  & 0.22 \\
46 & 4.2 &    2.44   & 0.65 &
85 & 6.6 &    3.30  & 1.41 &
89 & 3.1 &    -0.62  &-0.12 \\
82 & 10.6 &   0.52  & 0.34 &
102 & 3.6 &    -1.08  &-0.25 &
95 & 4.2 &    0.46  & 0.12 \\
105 & 1.0 &    -0.23  &-0.01 &
117 & 1.3 &    -1.20  &-0.10 &
122 & 4.0 &    -0.28  &-0.07 \\
\end{tabular}
\label{pmode.tbl}
\end{table*}

\enddocument